# Predicting origin-destination ride-sourcing demand with a spatio-temporal encoder-decoder residual multi-graph convolutional network

Jintao Ke[a], Xiaoran Qin[a*], Hai Yang[a], Zhengfei Zheng[a], Zheng Zhu[a], Jieping Ye[b,c]

[a] Department of Civil and Environmental Engineering, The Hong Kong University of Science and Technology, Clear Water Bay, Kowloon, Hong Kong, China
[b] Department of Computational Medicine and Bioinformatics, University of Michigan, Ann Arbor, United States
[c] AI Labs, Didi Chuxing, Beijing, China

**Abstract**

With the rapid development of mobile-internet technologies, on-demand ride-sourcing services have become increasingly popular and largely reshaped the way people travel. Demand prediction is one of the most fundamental components in supply-demand management systems of ride-sourcing platforms. With accurate short-term prediction for origin-destination (OD) demand, the platforms make precise and timely decisions on real-time matching, idle vehicle reallocations and ride-sharing vehicle routing, etc. Compared to zone-based demand prediction that has been examined by many previous studies, OD-based demand prediction is more challenging. This is mainly due to the complicated spatial and temporal dependencies among demand of different OD pairs. To overcome this challenge, we propose the *Spatio-Temporal Encoder-Decoder Residual Multi-Graph Convolutional network* (ST-ED-RMGC), a novel deep learning model for predicting ride-sourcing demand of various OD pairs. Firstly, the model constructs OD graphs, which utilize adjacent matrices to characterize the non-Euclidean pair-wise geographical and semantic correlations among different OD pairs. Secondly, based on the constructed graphs, a residual multi-graph convolutional (RMGC) network is designed to encode the contextual-aware spatial dependencies, and a long-short term memory (LSTM) network is used to encode the temporal dependencies, into a dense vector space. Finally, we reuse the RMGC networks to decode the compressed vector back to OD graphs and predict the future OD demand. Through extensive experiments on the for-hire-vehicles datasets in Manhattan, New York City, we show that our proposed deep learning framework outperforms the state-of-arts by a significant margin.

* Corresponding author: Xiaoran Qin, e-mail: xqinad@connect.ust.hk

## 1. Introduction

Ride-sourcing service provided by transportation network companies (TNCs) such as Uber, Lyft and DiDi, has experienced rapid growth since its emergence in 2009. It is reported that Uber has expanded its business to 700 cities and 24 countries around the world, while DiDi is providing service for over 25 million trips on each day in 400 cities in China. With their 24-hour-a-day availability and capacity to serve door-to-door on-demand requests, ride-sourcing service is becoming an important and indispensable component in urban transportation systems. The major challenges in the operational management of ride-sourcing services are how to address supply-demand imbalance across space and time, and how to satisfy as many passenger requests as possible with limited vehicle fleet size. To address these issues, prior studies have proposed a series of approaches, including surge pricing that suppresses passenger demand in peak-hours (Zha et al., 2016; Xu et al., 2017; Zha et al., 2017), idle vehicle reallocations that move idle vehicles from regions with excessive supply to regions with excessive demand (Lin et al., 2018), efficient order dispatching strategies (Xu et al., 2018), rush hour supply management (Su et al. 2019) and shared ride-sourcing services that allow one vehicle to serve two or more passengers in each ride to improve vehicle usage (Ke et al., 2019a, b, c; Li et al. 2019; Dong et al. 2018; Liu et al. 2017; Nourinejad et al. 2016), etc.

Many of these strategies rely on accurate real-time demand forecasting, especially the real-time origin-destination (OD) demand forecasting that not only predicts the potential demand originating from one region but also their destinations. For example, being aware of that many passengers will raise ride requests from region A to region B, the platform can reallocate idle vehicles to region A in advance, such that passengers will not experience a long waiting time. More importantly, OD demand prediction plays a critical role in the operations of shared ride-sourcing services. To attract passengers to use shared ride-sourcing services, the platform usually implements a discounted fare for shared ride services compared with normal solo ride services; however, passengers who opt for shared rides may experience extra trip time caused by vehicles' detour to pick-up and drop-off other passengers. Generally, as the platform decides to dispatch a vehicle to pick-up the first passenger, it does not know whether the vehicle can easily pick-up a second passenger en-route. This uncertainty makes it difficult for the platform to determine the upfront trip fares and dispatching strategies (whether and when to dispatch an en-route vehicle to pick-up a second passenger). An accurate OD demand forecasting can alleviate these uncertainties, and help the platform estimate the probability that an en-route vehicle meets a second passenger, which can further guide the upfront pricing and matching decisions. As the market share of shared ride-sourcing services increases rapidly, for example, Lyft is said to have 50 percent of its rides being shared by 2022 (Schaller, 2018), the need for an accurate prediction of OD demand becomes more and more urgent.

Most of the existing studies focus on the prediction of passenger demand originating from each region or zone (Ke et al., 2017; Zhang et al., 2017; Yao et al., 2018a; Zhang et al., 2019). However, there are only a few attempts (Liu et al., 2019; Wang et al., 2019) towards predictions of OD passenger demand. The possible reason is that OD-based demand forecasting is much more challenging than zone-based demand forecasting. One of the major challenges is how to capture the spatial-temporal dependencies between each two OD pairs. On a complex and irregular network, passenger demand in different OD pairs can be correlated with each other both geographically and semantically. For example, as shown in Figure 1, the passenger demand originating from a subway station S to commercial zone A and the passenger demand from the station S to commercial zone B can be positively correlated during some time intervals such as morning rush hours, since they are both determined by the outflow of the subway station. This indicates that the prediction of demand for a specific OD pair can benefit from the information of OD pairs with nearby origins or destinations. Moreover, an OD pair may have common demand patterns with a geographically distant OD pair, as they share similar functionalities. As the example in Figure 1, an OD pair from residential zone C to commercial zone B is similar to an OD pair from residential zone D to commercial zone A semantically, even though A and D are geographically far away from B and C, respectively. In spite of their importance, the spatial-temporal dependencies among different OD pairs are not well modeled and characterized by the existing methods.

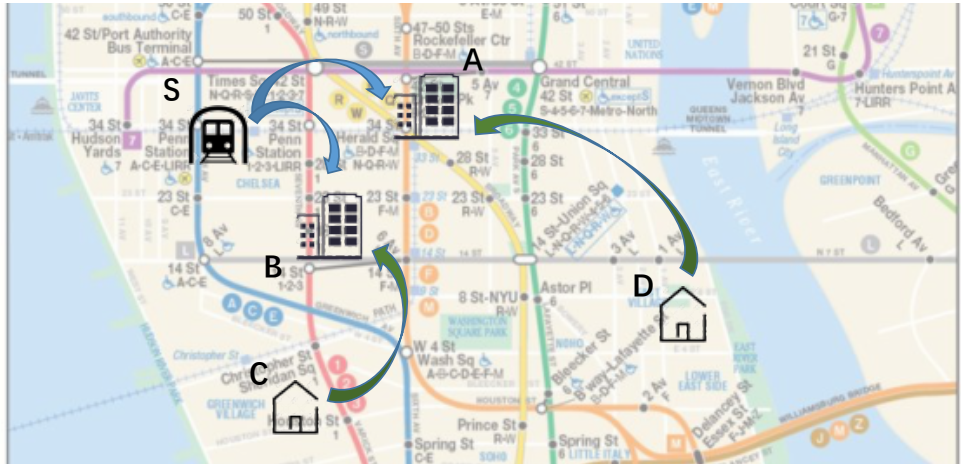

**Figure 1.** An example of different correlations among OD pairs

To overcome the aforementioned challenge, this paper proposes a novel deep learning framework named *Spatial-Temporal Encoder-Decoder Residual Multi-Graph Convolutional network* (ST-ED-RMGC) to simultaneously predict ride-sourcing passenger demand in various OD pairs. First, we construct multiple OD graphs, in which each OD pair is viewed as a node, and the adjacent matrices of nodes are established to represent different aspects of relationships among the OD pairs, such as neighborhood, distance, functional similarity, and historical demand correlations, and so on. Second, to capture both spatial and temporal correlations, we use a residual multi-graph convolutional (RMGC) network to capture the spatial correlations among OD pairs in different time intervals, and a regular long-short term memory (LSTM) network to characterize the temporal correlations of each OD pair itself. Being aware of the different input/output formats of the RMGC network (with graphs as inputs and outputs) and LSTM network (with time series as inputs and outputs), we encode the outputs of these two networks into a dense latent vector space. Then the RMGC network is reused to decode the latent vector back to the OD graph to predict future OD demand. The main contributions of this paper are summarized as follows:

- We characterize the pair-wise relationships between different OD pairs by constructing multiple OD graphs, including origin- and destination-based neighborhood relationship graphs, origin- and destination-based functional similarity graphs, origin- and destination-based distance graphs, and mobility pattern correlation graph.
- A novel deep learning model is designed to model both the spatial dependencies across different OD pairs and the temporal dependencies of the OD pairs themselves. A well-designed encoder-decoder structure is proposed to learn spatial and temporal features in an end-to-end learning framework.
- Evaluated by the datasets of for-hire-vehicles (namely, vehicles providing ride-sourcing services) in Manhattan, New York City, the proposed model significantly outperforms the benchmark algorithms.

The rest of the paper is organized as follows: Section 2 summarizes the recent studies related to ride-sourcing demand forecasting; Section 3 gives a clear definition of the research problem and spatio-temporal features used as inputs. Section 4 describes the proposed ST-ED-RMGC model from an overall architecture to its detailed components. Section 5 presents the dataset and the experimental results, and Section 6 concludes the paper and outlooks future studies.

## 2. Related Work

*Demand forecasting*

In transportation systems, real-time prediction of passenger demand or traffic states are fundamental tasks that can guide further system planning and operational strategy design. In the literature, many efforts have been directed towards prediction of traffic flow (Zhang et al. 2019; Yu et al. 2019; Wang et al. 2018; Zhang et al. 2019; Li et al. 2017; Wu et al. 2018; Zhu et al., 2016; Zhu et al., 2018; Zhu et al., 2019; Guo et al. 2019) as well as bike flow (Chai et al. 2018; Lin et al. 2018). Recently, the emergence of ride-sourcing services has generated tremendous mobility data, which makes the real-time prediction of ride-sourcing demand possible. Many of these prior studies (Ke et al., 2017; Zhang et al., 2017; Yao et al., 2018a; Zhang et al., 2019; Yao et al. 2018b) examined zone-based demand prediction problems. They partitioned a city into various squares with horizontal and vertical lines, and predict the near future (i.e., from 10 minutes to 2 hours in advance) passenger demand originating from each square. There are also some studies (Ke et al., 2018) which used regular hexagons as the basic grids for conducting predictions, because they found that hexagons resembles a circle and can better characterize the inflow and outflow between neighboring grids. In this way, researchers can treat the input and output data of ride-sourcing demand as images, and adopt some deep learning techniques that are widely used in image recognitions, such as convolutional neural networks (CNNs) and LSTM networks, etc.

However, CNNs only capture the local spatial correlations in a geographical manner, but fails to models the semantic correlations between two zones which are far away from each other in geographical but share common functionalities. Moreover, CNNs are not well adapted to predictions on irregular zones (such as administrative regions, Zip-code regions), which do not have an image-like data structure. To tackle this problem, some recent researches introduced graph convolutional neural network (GCN) and graph embedding techniques into ride-sourcing demand predictions. Yao et al. (2018a) proposed a deep multi-view spatial-temporal network, with three major views: spatial view, temporal view and semantic view. The semantic view uses graph embedding techniques to incorporate the information of functionalities of various zones into the framework. Sun et al. (2019) proposed a multi-view graph convolutional network to predict inflows and outflows in irregular regions. Geng et al. (2019a) proposed a ST-MGCN model to forecast zone-based ride-sourcing demand. The proposed model first characterizes the non-Euclidean relationship among zones by designing various adjacent matrices, and then applies recurrent neural networks to learn the temporal correlations. Geng et al. (2019b) further utilize grouped GCN in lower layer and multi-linear relationship GCN in higher layer to learn more generalized features.

While there is a large stream of literatures on zone-based demand prediction, there are only a few primary studies on OD-based demand prediction. Liu et al. (2019) proposed a contextualized spatial-temporal network to predict the OD taxi demand in New York. They partitioned the city into squares and used a 3D matrix (with two dimensions of height and width of the city map and one dimension of the total number of squares) to encode the origin-destination information, such that various types of CNNs can be utilized to capture the spatial dependences among the square-shape regions. Wang et al. (2019) estimated the OD matrix with a grid-embedding based multi-task learning. The grid embedding module was used to obtain the pre-weighted features by modelling the spatial mobility patterns of passengers, which were then fed into the multi-task learning module to predict future OD demand. Xiong et al. (2019) fused line GCN and Kalman filter to predict the OD demand of the whole traffic network. However, they treated the origin and destination grids separately and learned the local features around the separate grids. The adjacent matrices did not take advantage of semantic information, but only contain information of the distances and flows between any two grids.

*Graph convolution*

CNNs have been widely adopted in the image processing field due to its outstanding performance in capturing local spatial correlations. However, one significant drawback of CNNs is that they require input and output data as matrices or tensors, which makes them hard to be adapted to many real-world problems that rely on arbitrary graphs with non-Euclidean correlations. To address this issue, a large number of GCNs have been proposed that re-define the convolution operators for graph data, since their emergence (Bruna et al. 2013). Some studies (Li et al. 2018; Levie et al. 2017; Kipf and Welling) employed spectral graph theory into GCNs to transfer information from original graph domain to frequency domain to capture the non-Euclidean relationships among vertices. Others directly perform convolution operations in the original graph domain by aggregating the features of neighboring nodes (Gao et al. 2018; Monti et al. 2017). The later type of methods can be computed in a batch of nodes instead of the whole graph to reduce computational complexity. To see a comprehensive review on GCNs, readers may refer to Wu et al. (2019). Nowadays, GCNs are popular in transportation research due to its outperformance in many applications, such as traffic flow prediction (Wu et al. 2018; Do et al. 2019), traffic speed and state prediction (Zhang et al. 2019), parking occupancy prediction (Yang et al. 2019), etc.

## 3. Research problem

In this section, we will clarify our research problem by giving a clear definition to the OD graph. Besides, the geographical, semantic and temporal features used in our paper are described.

### 3.1 OD Graph

As mentioned above, many previous studies divided the examined space into various regular grids, such as squares and hexagons. These segmentations enable the use of standard machine learning algorithms, such as (CNN), but cannot well represent the administrative and functional properties of the regions. In this paper, we partition the examined city, Manhattan in New York City, into various irregular zones, according to the administrative zip codes (as shown in Figure 2 (a)). Each day is uniformly divided into several intervals (for example, 24 hours). The target of this paper is to predict the quantity of requested orders in various OD pairs simultaneously in each time interval.

Unlike the conventional traffic network graph with each vertex representing an interaction or zone, in this paper, we construct a tailored OD graph $G = (V, E, \boldsymbol{A})$ in which each vertex of the graph refers to an OD pair, as shown in Figure 2 (b). $V$ is the set of OD pairs to be predicted, $N = |V|$ is the number of OD pairs in each time interval, $E$ denotes the set of edges, and $\boldsymbol{A} \in \mathbb{R}^{N \times N}$ defines the adjacent matrixes with their entries representing the connections between vertices (i.e. OD pairs). Note that the OD graph is fully connected, indicating that there exists an edge connecting any two OD pairs, although their connections could be weak due to far geographical and semantic distances.

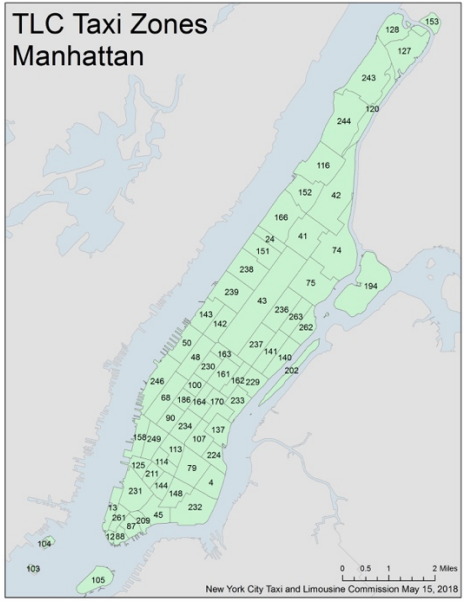 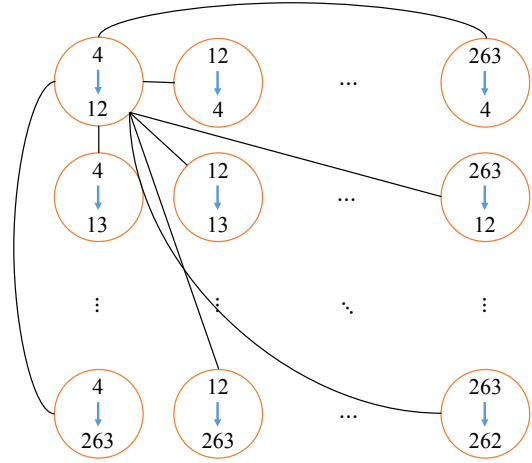

(a) Irregular zones based on zip codes

(b) Sample fully connected graph with OD pair as nodes (e.g. the 1st node is the OD pair with origin No. 4 and destination No. 12, connect to every node else)

**Figure 2.** Zone division and fully connected graph

### 3.2 Research problem and features

Let $x_i^{(d,t)}$ denote the passenger demand (quantity of requested orders) in the $i$th OD pair at time interval $t$ of day $d$, where $i \in V$ (the set of OD pairs), $x_i^{(d,t)} \in \mathbb{R}^+$. Let $X^{(d,t)}$ denote the passenger demand in all OD pairs at time interval $t$ of day $d$. To predict $X^{(d,t)}$, all the OD demand prior to time interval $t$ of day $d$ can be used as features. However, feeding all of the historical OD demand into the model is unnecessary and infeasible due to the limitations of computational resources. As pointed out by Zhang et al. (2017), there are two major types of temporal dependencies: tendency (demand is affected by the historical demand in the past few intervals) and periodicity (demand repeats similar patterns over days and over weeks). With this knowledge, we extract the following historical observations as the features:

1) Tendency-based features: the demand in the OD graph at the last two time intervals, i.e. $X^{(d,t-1)}$ and $X^{(d,t-2)}$;

2) Periodicity-based features over day: the demand in the OD graph at the same time interval in the previous day, i.e. $X^{(d-1,t)}$;

3) Periodicity-based features over week: the demand in the OD graph at the same time interval in the same day of last week, i.e. $X^{(d-7,t)}$.

Then the OD ride-sourcing demand prediction problem can be formulated by,

**Problem 1**: To learn a function $f(\cdot)$: $\mathbb{R}^{N \times T} \to \mathbb{R}^N$ that maps the historical demand of all OD pairs on an OD graph to the demand of all OD pairs on the same OD graph in the next time interval:

$$\widetilde{X}^{(d,t)} = \left[X^{(d-7,t)}, X^{(d-1,t)}, X^{(d,t-2)}, X^{(d,t-1)}\right] \xrightarrow{f} X^{(d,t)} \tag{1}$$

## 4. The Proposed ST-ED-RMGC Model

### 4.1 Overview of model framework

Figure 3 briefly demonstrates the architecture of the proposed ST-ED-RMGC model, which uses an encoder-decoder framework. There are two encoders: one spatial encoder and one temporal encoder. The spatial encoder utilizes several RMGCs to model the spatial correlations between the OD pairs from different aspects (including geographical distances, functionality similarities, and mobility pattern correlations). The temporal encoder proposes a spatial LSTM model to learn the temporal dependencies of each OD pair. To fuse the spatial and temporal models in one end-to-end learning framework, we flatten the outputs of the two decoders into two dense latten vectors, which are then concatenated. Finally, in the decoder part, several RMGC networks are reused to transform the compressed vector back to an OD graph, which is used to predict the target OD demand.

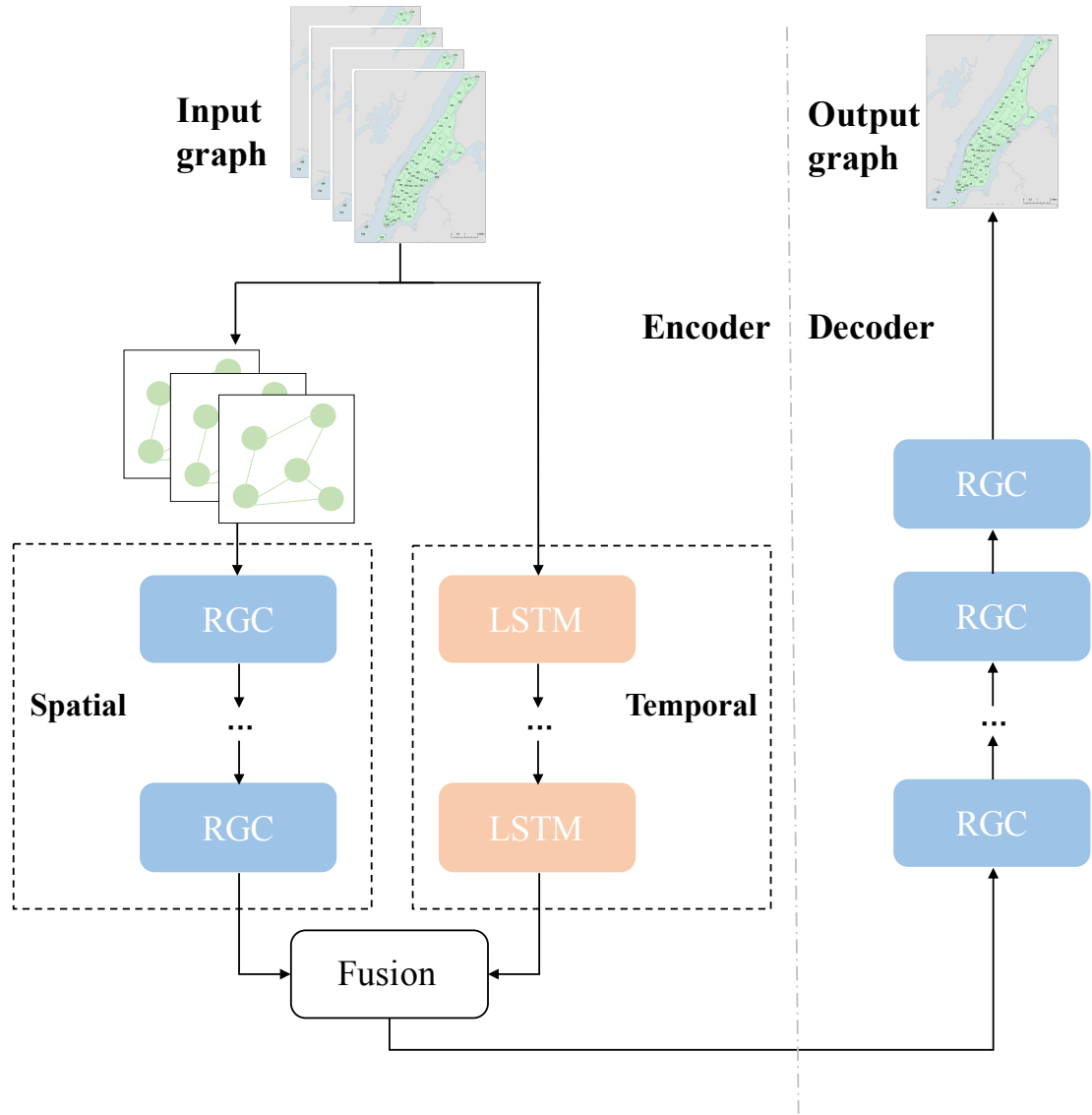

**Figure 3.** Framework of ST-ED-RMGC Model

**4.2 Detailed methodologies**

This section presents the detailed methodologies of different modules in the framework. We will start with designs of the multiple graphs for capturing various types of spatial relationships among the OD pairs, and then present the technical details of the RMGC module. Besides, we will give a brief introduction to the spatial LSTM module, which is slightly different from the standard LSTM network. Finally, we describe the encoder-decoder architecture that combines the abovementioned modules in one end-to-end learning framework.

*4.2.1 Modelling spatial correlations between OD pairs*

This subsection models the multi-graphs and defines the corresponding adjacent matrixes. The dependency among OD pairs depicts the similarity or connectivity of regions in OD demand prediction. We propose four types of graph to model the spatial correlations between OD pairs: (1) neighborhood relationship graphs $G_n(V, E, A_n)$, $A_n \in \mathbb{R}^{N \times N}$; (2) functional similarity graphs $G_f(V, E, A_f)$, $A_f \in \mathbb{R}^{N \times N}$; (3) centroid distance graphs $G_d(V, E, A_d)$, $A_d \in \mathbb{R}^{N \times N}$; and (4) mobility pattern correlation graph $G_c(V, E, A_c)$, $A_c \in \mathbb{R}^{N \times N}$.

(1) *Neighborhood relationship graphs*. As mentioned in Section 1, demand in OD pairs with nearby origin or destination are more likely to have similar patterns. We define two adjacent matrices to indicate whether two OD pairs have neighboring origins or destinations, respectively:

$$[A_n^O]_{i,j} = \begin{cases} 1, & \text{if origin of OD pair i and OD pair j are adjacent} \\ 0, & \text{otherwise} \end{cases}, \forall i,j \epsilon V$$

$$[A_n^D]_{i,j} = \begin{cases} 1, & \text{if destinations of OD pair i and OD pair j are adjacent} \\ 0, & \text{otherwise} \end{cases}, \forall i,j \epsilon V \quad (2)$$

where $[A_n^O]_{i,j}$ refers to the element in the $i$th row and $j$th column of matrix $A_n^O$.

(2) *Functional similarity graphs*. Different regions in a city may have different functionalities or land-use properties. Some are commercial regions with many shopping malls and restaurants, some are tourist areas with a park, and some are residential zones full of departments and houses. In this paper, we use the land-use properties provided by Smart Location Database (please refer to the details in Section 5.1) to define the functionalities of zones in Manhattan. The properties we select involve households without owning an automobile, house density, population density, employment density, road density as well as the average meters to nearest transit, which are highly related to the land-use type and travel mode choice. Note that the administrative zones have heterogeneous areas, and thus we divide all these measures by the area of the corresponding zone. Let $F_i^O$ and $F_i^D$ denote the vector of the functionalities of the origin zone and destination zone of OD pair $i$, then two functional similarity adjacent matrices can be constructed as follows:

$$[A_f^O]_{i,j} = \left[ \sqrt{(F_i^O - F_j^O)(F_i^O - F_j^O)^T} \right]^{-1}, \forall i,j \epsilon V \quad (3)$$

$$[A_f^D]_{i,j} = \left[ \sqrt{(F_i^D - F_j^D)(F_i^D - F_j^D)^T} \right]^{-1}, \forall i,j \epsilon V \quad (4)$$

We can see that, as the vectors of functionalities $F_i^O, F_j^O$ of two OD pairs $i, j$ become closer to each other, their functional similarity gets larger.

(3) *Centroid distance graphs*. Due to the irregular zones with different sizes, we further introduce two centroid distance graphs to represent the geographical relationships between OD pairs. The adjacent matrices in the origin-based (distance between origin centroids of each two OD pairs) and destination-based centroid distance (distance between destination centroids of each two OD pairs) graphs are defined by the inverse of the straight-line distance between the centroids of the zones (the shorter the distance, the stronger the relationship), shown as follows:

$$[A_d^O]_{i,j} = [haversine(lng_i^O, lat_i^O, lng_j^O, lat_j^O)]^{-1}, \forall i,j \epsilon V \qquad (5)$$

$$[A_d^D]_{i,j} = [haversine(lng_i^D, lat_i^D, lng_j^D, lat_j^D)]^{-1}, \forall i,j \epsilon V \qquad (6)$$

where $lng_i^O, lat_i^O, lng_j^O, lat_j^O$ are the longitudes and latitudes of the origins of OD pairs $i$, $j$, while $lng_i^D, lat_i^D, lng_j^D, lat_j^D$ are the longitudes and latitudes of the destinations of OD pairs $i, j$, respectively. The function $haversine(\cdot)$ measures the straight-line distance between two locations on earth.

(4) *Mobility pattern correlation graphs*. It is intuitive that OD pairs with analogous mobility patterns (historical demand trends) share some common characteristics, and thus can guide predictions for each other. Let $Q_i$, be a vector recording the historical demand (over multiple months) of OD pair $i$. Then the adjacent matrix of the mobility pattern correlation graph is formulated by,

$$[A_c]_{i,j} = \frac{Cov(Q_i, Q_j)}{\sqrt{var(Q_i)var(Q_j)}}, \forall i,j \epsilon V \qquad (7)$$

where $Cov(\cdot,\cdot)$ calculates the covariance between two vectors, while $var(\cdot)$ calculates the variance of one vector. Figure 4 provides an illustration of the adjacent matrixes.

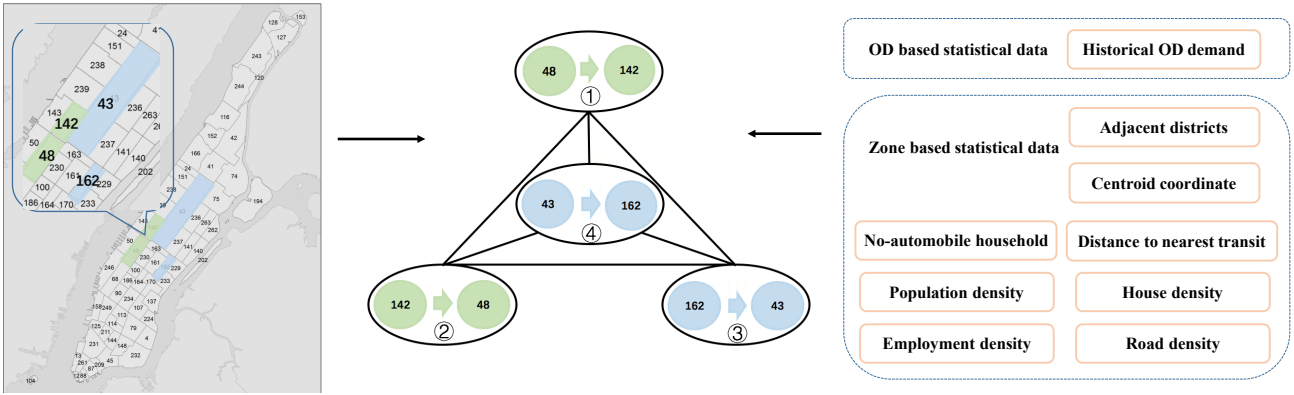

(a) Graph construction and input information

## Origin neighborhood relationship

$$A_n^O = \begin{array}{c} \\ ① \\ ② \\ ③ \\ ④ \end{array} \begin{bmatrix} ① & ② & ③ & ④ \\ 0 & 1 & 0 & 0 \\ 1 & 0 & 0 & 1 \\ 0 & 0 & 0 & 0 \\ 0 & 1 & 0 & 0 \end{bmatrix}$$

## Destination neighborhood relationship

$$A_n^D = \begin{array}{c} \\ ① \\ ② \\ ③ \\ ④ \end{array} \begin{bmatrix} ① & ② & ③ & ④ \\ 0 & 1 & 1 & 0 \\ 1 & 0 & 0 & 0 \\ 1 & 0 & 0 & 0 \\ 0 & 0 & 0 & 0 \end{bmatrix}$$

(b) Neighborhood relationship adjacent matrix

## OD order quantity correlation

$$A_c = \begin{array}{c} \\ ① \\ ② \\ ③ \\ ④ \end{array} \begin{bmatrix} ① & ② & ③ & ④ \\ 0 & cor(Q_1,Q_2) & cor(Q_1,Q_3) & cor(Q_1,Q_4) \\ cor(Q_1,Q_2) & 0 & cor(Q_2,Q_3) & cor(Q_2,Q_4) \\ cor(Q_1,Q_3) & cor(Q_2,Q_3) & 0 & cor(Q_3,Q_4) \\ cor(Q_1,Q_4) & cor(Q_2,Q_4) & cor(Q_3,Q_4) & 0 \end{bmatrix}$$

(c) Mobility pattern correlation adjacent matrix

## Origin centroid distance

$$A_d^O = \begin{array}{c} \\ ① \\ ② \\ ③ \\ ④ \end{array} \begin{bmatrix} ① & ② & ③ & ④ \\ 0 & d_{48,142}^{-1} & d_{48,162}^{-1} & d_{48,43}^{-1} \\ d_{142,48}^{-1} & 0 & d_{142,162}^{-1} & d_{142,43}^{-1} \\ d_{162,48}^{-1} & d_{162,142}^{-1} & 0 & d_{162,43}^{-1} \\ d_{43,48}^{-1} & d_{43,142}^{-1} & d_{43,162}^{-1} & 0 \end{bmatrix}$$

## Destination centroid distance

$$A_d^D = \begin{array}{c} \\ ① \\ ② \\ ③ \\ ④ \end{array} \begin{bmatrix} ① & ② & ③ & ④ \\ 0 & d_{142,48}^{-1} & d_{142,43}^{-1} & d_{142,162}^{-1} \\ d_{48,142}^{-1} & 0 & d_{48,43}^{-1} & d_{48,162}^{-1} \\ d_{43,142}^{-1} & d_{43,48}^{-1} & 0 & d_{43,162}^{-1} \\ d_{162,142}^{-1} & d_{162,48}^{-1} & d_{162,43}^{-1} & 0 \end{bmatrix}$$

(d) Centroid distance adjacent matrix

## Origin functional similarity

$$A_f^O = \begin{array}{c} \\ ① \\ ② \\ ③ \\ ④ \end{array} \begin{bmatrix} ① & ② & ③ & ④ \\ 0 & E_{48,142}^{-1} & E_{48,162}^{-1} & E_{48,43}^{-1} \\ E_{142,48}^{-1} & 0 & E_{142,162}^{-1} & E_{142,43}^{-1} \\ E_{162,48}^{-1} & E_{162,142}^{-1} & 0 & E_{162,43}^{-1} \\ E_{43,48}^{-1} & E_{43,142}^{-1} & E_{43,162}^{-1} & 0 \end{bmatrix}$$

## Destination functional similarity

$$A_f^D = \begin{array}{c} \\ ① \\ ② \\ ③ \\ ④ \end{array} \begin{bmatrix} ① & ② & ③ & ④ \\ 0 & E_{142,48}^{-1} & E_{142,43}^{-1} & E_{142,162}^{-1} \\ E_{48,142}^{-1} & 0 & E_{48,43}^{-1} & E_{48,162}^{-1} \\ E_{43,142}^{-1} & E_{43,48}^{-1} & 0 & E_{43,162}^{-1} \\ E_{162,142}^{-1} & E_{162,48}^{-1} & E_{162,43}^{-1} & 0 \end{bmatrix}$$

(e) Functional similarity adjacent matrix

**Figure 4.** Adjacent matrixes processing

### 4.2.2 Residual Multi-Graph Convolutional (RMGC) Network

Now we introduce the RMGC network that combines a multi-graph convolutional network and a residual module, to capture the spatial correlations between OD pairs. Recently, GCNs have become increasingly popular and been widely used for many tasks, including social network analysis, abnormal detection detections on graphs, graph embedding and traffic forecasting, etc. There are majorly two types of GCNs: spatial-based and spectral-based. Spatial-based networks use local graph convolution units to extract feature information from neighboring vertices, while spectral-based approaches (Defferrard et al., 2016) introduce spectral filters to define graph convolutions. In this paper, we focus on the spectral-based methods, which in essence map the original graph signals (raw features on a graph) to a parameterized Fourier domain. However, training the parameters are computationally expensive. To address this issue, Defferrard et al. (2016) introduced a Chebyshev polynomial expansion (up to $K$th order) to obtain an efficient approximation, and Kipf and Welling (2016) further simplified the spectral filter to a Chebyshev polynomial of order $K = 1$. The later model is also called 1stChebNet and has the following form:

$$\boldsymbol{H}_{l+1} = \sigma(\widehat{\boldsymbol{A}}\boldsymbol{H}_l\boldsymbol{W}_l) = \sigma(\boldsymbol{D}^{-1/2}\widetilde{\boldsymbol{A}}\boldsymbol{D}^{-1/2}\boldsymbol{H}_l\boldsymbol{W}_l) \qquad (8)$$

where $\boldsymbol{H}_{l+1}$ and $\boldsymbol{H}_l$ are the activations in the $l$th and $l+1$th hidden layer; $\sigma$ is the activation function, which can be $Relu(\cdot)$ or $Linear(\cdot)$ or $Tahn(\cdot)$; $W_l$ denotes the trainable weight matrix connecting $l$th and $l+1$th hidden layer. $\widetilde{\boldsymbol{A}} = \boldsymbol{A} + \boldsymbol{I}$ is the adjacent matrix added by self-connections for the purpose of maintaining the information of the node itself in convolution (Chai, D. et al. 2018). $\boldsymbol{D}$ is the degree matrix, in which $\boldsymbol{D}_{ii} = \sum_j \widetilde{\boldsymbol{A}}_{ij}$ and $\boldsymbol{W}_l$ is a matrix of trainable weights in the $l$th graph convolutional layer.

The transformation function in Eq. (8) describes the mapping from a hidden layer to the next hidden layer through one graph with a single adjacent matrix. To enable the learning from multiple graphs, we design a training architecture shown in Figure 5. Suppose we have $K$ adjacent matrixes, and let $\widehat{\boldsymbol{A}_k} \in \mathbb{R}^{N \times N}$ denote the $k$th adjacent matrix, where $k\epsilon\{1, \dots, K\}$. In each training batch with a batch size $B$, we first duplicate each adjacent matrices by $B$ times and then concatenate them into a tensor $\ddot{\boldsymbol{A}} \in \mathbb{R}^{B \times (N*K) \times N}$. Suppose the dimensions of the input and output features are $F$ and $O$ respectively, then we have the input tensor $\boldsymbol{H}_l \in \mathbb{R}^{B \times N \times F}$, the learnable weight matrix $\boldsymbol{W} \in \mathbb{R}^{(K*F) \times O}$, and the output tensor $\boldsymbol{H}_{l+1} \in \mathbb{R}^{B \times N \times O}$. To map the input tensor $\boldsymbol{H}_l$ to the output tensor $\boldsymbol{H}_{l+1}$, we first conduct a batch dot production between $\ddot{\boldsymbol{A}}$ and $\boldsymbol{H}_l$, which generates a tensor $\boldsymbol{M} \in \mathbb{R}^{B \times (N*K) \times F}$. Then the generated tensor $\boldsymbol{M}$ is reshaped to a new tensor $\widetilde{\boldsymbol{M}} \in \mathbb{R}^{B \times N \times (F*K)}$, and finally the batch dot product of $\widetilde{\boldsymbol{M}}$ and $\boldsymbol{W}$ produces the output tensor $\boldsymbol{H}_{l+1} \in \mathbb{R}^{B \times N \times O}$.

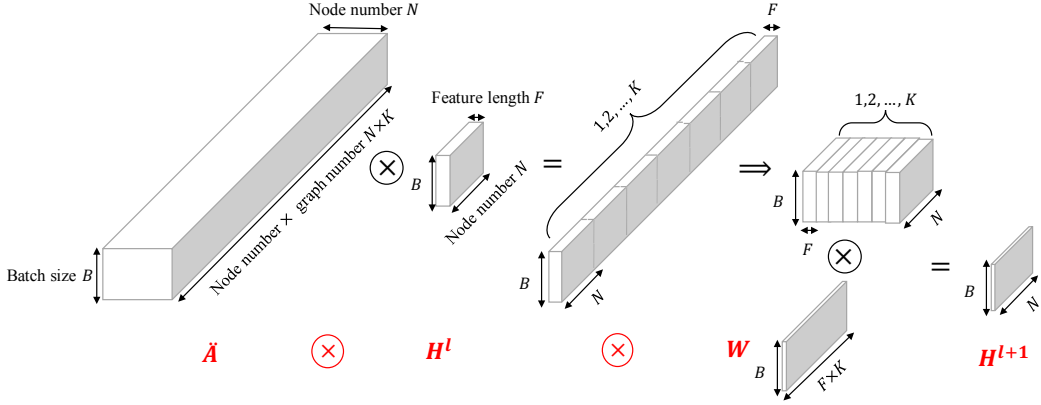

**Figure 5.** Architecture of one MGC layer

To train the networks in a deep neural network structure without suffering from gradient explosion, now we develop two multi-graph convolutional network (MGC) based residual blocks, the identity block and convolutional block. Residual learning (He et al., 2015) is a powerful tool that allows trainings of super deep networks, and is widely used in many traditional convolutional neural network structures. The basic residual unit (identity block) is formulated by,

$$\boldsymbol{H}_{l+1} = \boldsymbol{H}_l + \sigma\big(\boldsymbol{D}^{-1/2}\widetilde{\boldsymbol{A}}\boldsymbol{D}^{-1/2}\boldsymbol{H}_l\boldsymbol{W}_l\big) \tag{9}$$

where the first term at the RHS refers to the shortcut path, and the second term represents the main path. It is noteworthy that a common practice is to stack multiple MGC layers in the main path. Moreover, in the case that requires different feature dimensions between the input tensor $\boldsymbol{H}_l$ and the output tensor $\boldsymbol{H}_{l+1}$, a MGC layer can be added to the shortcut path to maintain the feature dimension consistency (as shown in Figure 6). This residual block is called the convolutional block.

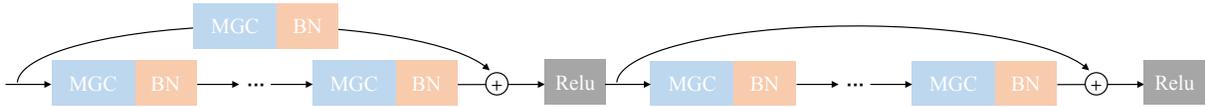

**Figure 6.** Convolutional block and identity block in RMGC

We stack multiple convolutional and identity RMGC blocks, and finally flatten the output into a latent vector with a one-dimensional feature:

$$\boldsymbol{L}_1 = \sigma\left(\boldsymbol{W}_1 \cdot Flatten\left(RMGC(\widetilde{\boldsymbol{X}}, \boldsymbol{W}_{RMGC})\right) + b_1\right) \tag{10}$$

where $RMGC(\cdot)$ refers to the transformation through multiple RMGC blocks with trainable weights $\boldsymbol{W}_{RMGC}$, $Flatten(\cdot)$ is an operator that flattens the outputs of the last RMGC layer, $\boldsymbol{W}_1$ and $b_1$ are trainable weights and bias in the latent layer. $L_1$ has a dimension of $B \times V_1$, where $B$ is the batch size and $V_1$ is the latent feature dimension of the RMGC encoder.

*4.2.3 Spatial LSTM*

LSTM, as a recurrent neural network, was born for capturing temporal dependencies, and thus are widely used in many traffic demand forecasting problems. In most of previous studies, it took the historical features of one zone or one road segment as inputs and predicts the demand of this zone or road segment in the next time interval. In our paper, in order to incorporate spatial and temporal models in one end-to-end learning framework in a more sensible way, we propose a spatial LSTM that learns features from all OD pairs and outputs the high-level information into a latent vector, instead of using various separate LSTMs for each OD pair. Recall that the input of features $\widetilde{X}$ has a shape of $B \times N \times T$, where $B$ is the batch size, $T$ is the number of sliced windows for extracting historical features ($T = 4$ as defined in Problem 1). The proposed spatial LSTM first reshapes $\widetilde{X}$ by transposing its second and third dimensions, leading to a new tensor with a shape of $B \times T \times N$. Then the LSTM treat the second dimension of $T$ as the time dimension, and the third dimension of $N$ as the feature dimension. In other words, the historical demand in all OD pairs are treated as features that are fed into one single LSTM. We stack multiple LSTM modules and flatten the output tensor to a latent vector:

$$L_2 = \sigma \left( W_2 \cdot Flatten \left( LSTM(\widetilde{X}, W_{LSTM}) \right) + b_2 \right) \qquad (11)$$

where $LSTM(\cdot)$ represents multiple LSTM layers with trainable weights $W_{LSTM}$, $Flatten(\cdot)$ flattens the outputs of the last LSTM layer, $W_2$ and $b_2$ are the weights and bias in the latent layer, $L_2$ has a dimension of $B \times V_2$, where $V_2$ is the latent feature dimension of the spatial LSTM encoder.

*4.2.4 Encoder fusion and decoder*

We simply fuse the outputs of the abovementioned two encoders by concatenating them together:

$$L = [L_1, L_2] \qquad (12)$$

where the latent vector $L$ contains both the spatial and temporal features of the historical demand. To predict the future OD demand, we first introduce an intermediate layer to expand the dimension of vector $L$ to a new vector with a shape of $B \times N$ ($N$ is the number of OD pairs), and then we reshape the new vector to a tensor with a shape of $B \times N \times 1$, which becomes a valid input format for RMGC. By stacking various RMGC modules, we can finally obtain the estimated demand of all OD pairs on an OD graph, i.e. $\widehat{X}^{(d,t)}$. Formally, the decoder architecture can be formulated by,

$$\widehat{X}^{(d,t)} = RMGC^d\big(Reshape(\sigma(W_3 \cdot L + b_3)), W_{RMGC-D}\big) \qquad (13)$$

where $RMGC^d(\cdot)$ are multiple RMGC modules with learnable weights $W_{RMGC-D}$, $W_3$ and $b_3$ are the trainable weights of the intermediate layer for dimension expansion, $Reshape(\cdot)$ is an operator that reshapes a vector into a tensor. Let $W, b$ be all the trainable weights and biases in the whole encoder-decoder architecture, we can train the weights and biases by solve the following optimization problem:

$$\min_{W,b} \sum_d \sum_t \|\widehat{X}^{(d,t)} - X^{(d,t)}\|_2^2 + \alpha \|W\|_2^2 \qquad (14)$$

where the first term minimizes the squared loss between the predicted OD demand and actual OD demand, and the second term is a L2-norm regularization term to avoid extremely complex models that may lead to over-fitting. The training algorithm of the model is demonstrated below.

Algorithm 1. ST-ED-RMGC training algorithm

---

**Algorithm 1** ST-ED-RMGC training algorithm
---
**Input:** OD pair number $i \in V$,
   Historical demand of all OD pairs $\{x_i^1, ..., x_i^T\}, \forall i \in V$,
   The graphs **G**: neighborhood relationship graphs $G_n(V, E, \mathbf{A}_n)$;
         functional similarity graphs $G_f(V, E, \mathbf{A}_f)$;
         centroid distance graphs $G_d(V, E, \mathbf{A}_d)$;
         mobility pattern correlation graph $G_c(V, E, \mathbf{A}_c)$
**Output:** ST-ED-RMGC with well-trained parameters **W**
   // Construct a set of input-output instances $\mathcal{D}$
   Initiallize a null set: $\mathcal{D} \leftarrow \emptyset$
   **for** time interval $t(1 \leq t \leq T)$ **do**
      Get temporal features of all OD pairs at each time interval: $\widetilde{\mathbf{X}}^{(d,t)} = [\mathbf{X}^{(d-7,t)}, \mathbf{X}^{(d-1,t)}, \mathbf{X}^{(d,t-2)}, \mathbf{X}^{(d,t-1)}]$
      // $\mathbf{X}^{(d,t)}$ is the prediction target at time t
      Put training sample into the dataset: $\mathcal{D} \leftarrow \mathcal{D} + (\widetilde{\mathbf{X}}^{(d,t)}, \mathbf{X}^{(d,t)})$
   **end for**
   Divide $\mathcal{D}$ into training and test datasets $\mathcal{D}_{train}, \mathcal{D}_{test}$

   // Training ST-ED-RMGC model
   Initialize the hidden status, all weights and bias parameters
   **for** $n = 0 \rightarrow$ number of epoch **do**
      Randomly select a batch of sample $\mathcal{D}_b$ from $\mathcal{D}_{train}$, where $b = 1, 2, ..., B$
      Assign temporal features of the batch to the training feature: $\mathbf{X} \leftarrow \widetilde{\mathbf{X}}_b^{(d,t)}$
      Assign prediction target of the batch to the ground truth: $\mathbf{y} \leftarrow \mathbf{X}_b^{(d,t)}$
      Calculate the simplified the spectral filter: $\hat{\mathbf{A}} \leftarrow \mathbf{D}^{-1/2}(\mathbf{A} + \mathbf{I})\mathbf{D}^{-1/2}$
      Reshape and assign the adjacent matrix: $\ddot{\mathbf{A}} \leftarrow \hat{\mathbf{A}}$ by duplicating for $B$ times, where $B$ is the batch size
      Optimize **W** by minimizing loss function Eq(14)
   **end for**
---

## 5. Experimental Results

To evaluate the performance of our ST-ED-RMGC model, we carry out an experiment utilizing the ride-sourcing data from New York City.

**5.1 Data Description**

The location map we used is from the Smart Location Database (SLD), a free data product and service provided by the U.S. EPA Smart Growth Program[†]. The Manhattan area is divided into districts according to administrative zip code, and each of them records the information including area, demographics, employment and transit. The database of ride-sourcing demand is the for-hire vehicle records from January 2018 to April 2018 collected by New York City Taxi & Limousine Commission. Each trip record is associated with attributes including: date, time, and taxi zone location ID (same with the ID in aforementioned location map) of pick-up and drop-off events[‡].

Next we use an example to demonstrate why OD demand prediction is challenging. Figure 7 illustrates the OD demand per hour of two representative OD pairs over one month. Firstly, we can see that, although the OD demand has some daily patterns or periodicities, the standard deviations of the OD demand at the same hour across different days are very high. This indicates that the daily patterns have strong fluctuations and uncertainties. Second, the daily patterns of different OD pairs are different from each other. As shown in the figure below, the demand patterns of two OD pairs with the same origins (zone No. 43) are quite different from each other. This implies that underlying spatial relationships between two OD pairs are hard to observe, even for OD pairs with the same origins or destinations. It makes the prediction of OD demand much more difficult than the zone-based demand prediction. Moreover, the OD matrix is very sparse and demand in most OD pairs is almost zero, which indeed increase the difficulty of precise prediction.

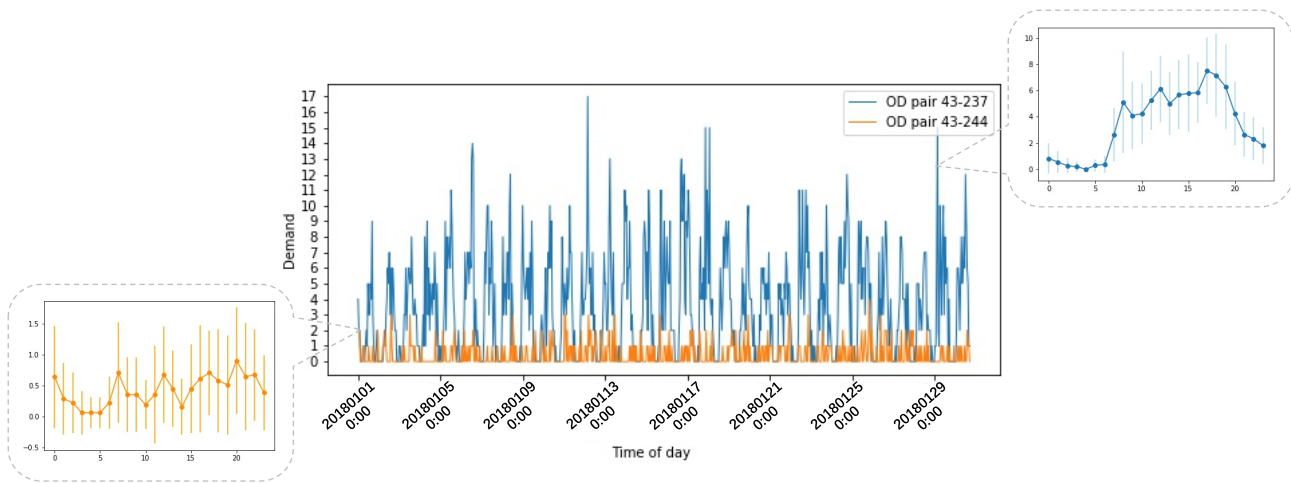

**Figure 7.** OD demand distribution

---

[†] https://www.epa.gov/smartgrowth/smart-location-mapping
[‡] https://www1.nyc.gov/site/tlc/about/tlc-trip-record-data.page

For data split, we use the data from January 8th 2018 to January 9th 2018 for training, and the data from November 1st 2018 to November 30th 2018 is used for testing.

**5.2 Model Setting**

In the encoder, we stack one RMGC convolutional block and one RMGC identity block. The main path of the RMGC convolutional block contains three MGC layers (with 32, 32, 128 hidden units), while the shortcut path contains one MGC layer (with 128 hidden units). The RMGC identity block has a main path containing three MGC layers (with 32, 32, 128 hidden units). On the other hand, the temporal encoder includes two LSTM layers with 128 and 64 hidden units respectively. The outputs of RMGC and LSTM encoders are flattened and then transformed into two latent vectors with dimensions of 900 and 100 respectively. The decoder uses one RMGC convolutional block and then one RMGC identity block, with the same settings as that in the encoder, which is then followed by a MGC layer that generates the estimated OD demand. All the activations in the hidden layers are Relu, while the activation in the output later is a linear function. The optimizer used in the model is Adam with a learning rate of 5e-5 and a decay of 1e-6. During the training phase, we set batch size to be 32.

**5.3 Models Comparison**

In this section, we compare the proposed ST-ED-RMGC with several traditional machine learning models and two graph convolutional networks, a MGC and a RMGC. The models are described below:

- HA: Historical average is one of the most fundamental statistical method of prediction. We use the average historical OD demand over the past four weeks.
- XGB: XGBoost is an implementation of gradient boosted decision trees designed for speed and performance (Chen et al. 2016). It dominates structured datasets on classification and regression predictive modeling problems.
- MLP: Multi-layer Perception is the basic neural network contains at least three layers, i.e. input layer, hidden layer and output layer. It is trained with the back-propagation.
- GBDT: Gradient Boosting Decision Tree is an iterative decision tree algorithm with multiple regression decision trees.
- RF: Random forest is the model that is trained by bootstrapped samples of each decision tree.
- LASSO: LASSO also takes historical data as input like normal regression, but with loss function considering least absolute shrinkage and selection operator.
- LSTM: Long short-term memory model is an artificial recurrent neural network with feedback connections. The gates and cells in the model are utilized to regulates and remember the information of a sequence.

- Spatial LSTM: a special type of LSTM that is described in Section 4.2.3.
- MGC network: a multi-graph convolutional network containing three MGC layers (with 32, 32, 128 hidden units).
- RMGC network: a residual multi-graph convolutional network with one RMGC convolutional block and one RMGC identity block (the same structure as the encoder).
- ST-ED-RMGC network: the parameters are presented in Section 5.2.

The parameters of all abovementioned models are fine-tuned. We assess the prediction error of the models by three metrics, RMSE (Root Mean Square Error), MAE (Mean Absolute Error) and MAPE (Mean Absolute Percentage Error). Due to the sparseness of OD demand, we only focus on the MAPE result of OD pairs whose demand is more than 1 unit (otherwise, the MAPE equals infinity if there is a zero demand). The performances of the models are shown in Table 1.

**Table 1**. Prediction error of baseline methods and proposed model with different parameters

| Method | RMSE | MAE | MAPE |
| --- | --- | --- | --- |
| HA | 8.42 | 5.40 | 0.75 |
| XGB | 5.20 | 3.59 | 0.48 |
| MLP | 5.20 | 3.59 | 0.49 |
| GBDT | 5.21 | 3.59 | 0.49 |
| RF | 5.61 | 3.87 | 0.51 |
| LASSO | 5.37 | 3.70 | 0.51 |
| LSTM | 5.25 | 3.64 | 0.51 |
| Spatial LSTM | 4.96 | 3.30 | 0.42 |
| MGC | 4.48 | 3.10 | 0.41 |
| RMGC | 4.46 | 3.07 | 0.40 |
| ST-ED-RMGC | 4.29 | 2.96 | 0.38 |

From the table, we can see that GCNs (MGC, RMGC and ST-ED-RMGC) significantly outperforms the traditional machine learning and deep learning methods such as MLP, XGBoost, LSTM. This implies that there exist strong spatial correlations among the OD pairs, which can be well captured by the proposed GCNs through the well-defined OD graphs and adjacent matrices. It can also be found that, although the residual units bring marginal gains to prediction accuracies (comparing RMGC with MGC), the encoder-decoder structure in ST-ED-RMGC further improve predictive performance on the basis of MGC. This indicates that the designed encoder-decoder framework is able to well combine the features learned from the spatial LSTM (temporal correlations) and RMGC (spatial correlations).

## 5.4 Prediction Results

To present the prediction result, we first examine the OD demand distribution of particular hours. We select the demand of the examined OD pairs that are distributed over 30 districts and build a 30 by 30 matrix to illustrate the prediction results, as shown in Figure 8. The left figures in each line show the ground truth values, the middle figures represent the predicted values, while the right figures demonstrate the relative prediction error (RE), of each OD pair. Two time intervals (20pm and 8am) are selected for illustration. It can be shown that the OD demand has strong imbalance across space and time. For example, people travel to district No. 2 and 6 in the morning and go back to district No. 16 in the evening. Our model well capture this human mobility patterns and thus makes precise predictions.

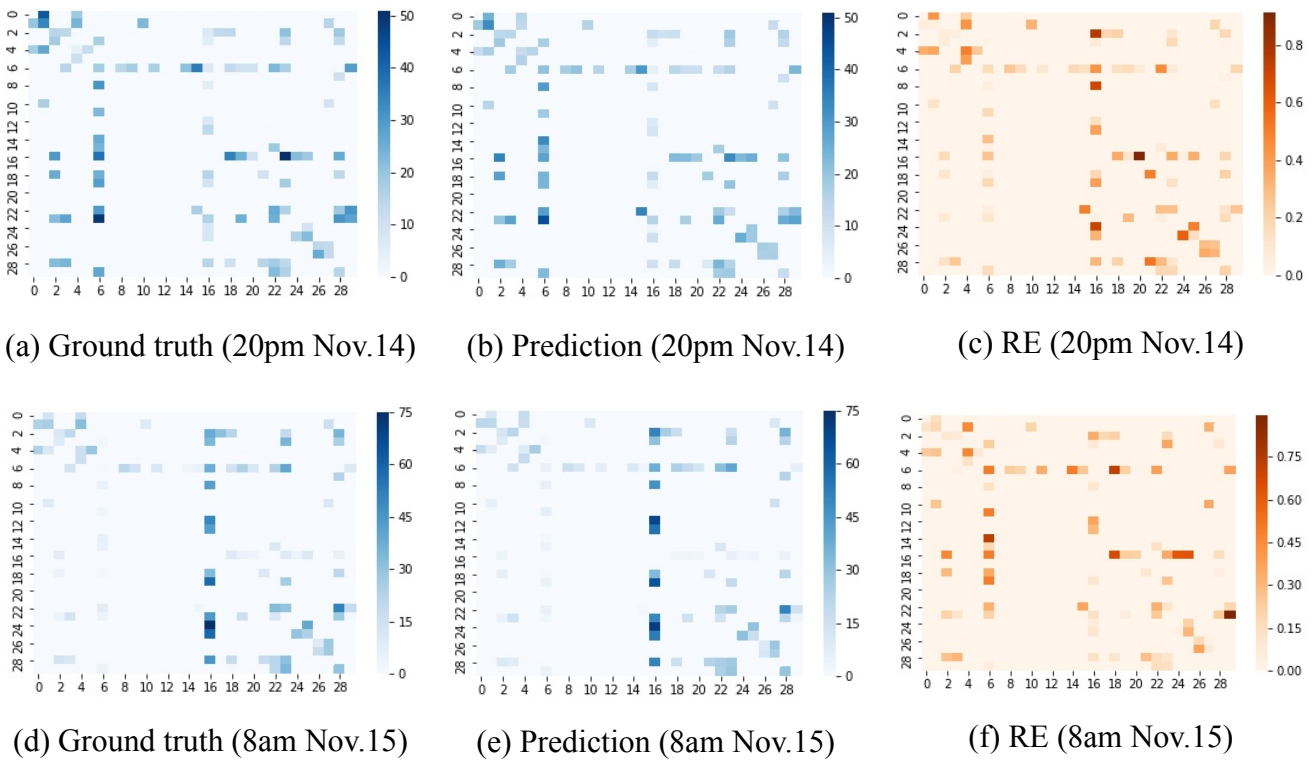

(a) Ground truth (20pm Nov.14)  (b) Prediction (20pm Nov.14)  (c) RE (20pm Nov.14)

(d) Ground truth (8am Nov.15)  (e) Prediction (8am Nov.15)  (f) RE (8am Nov.15)

**Figure 8.** The prediction result of OD distribution

To get a close look at the result, we select two OD pairs with high variance and different maximum demand volume. Figure 9 plots their two-week trends of the ground truth values, predicted values of one of the baseline (LASSO) and predicted values of the ST-ED-RMGC. It can be shown that the temporal patterns of OD demands are different across different days. In some days, there are two peaks; in other days, there is only one peak or a strong peak along with weak peak. Clearly, the ST-ED-RMGC can better captures the temporal fluctuations across different days, while the baseline is easily over-reacting or under-reacting to the unstable oscillations.

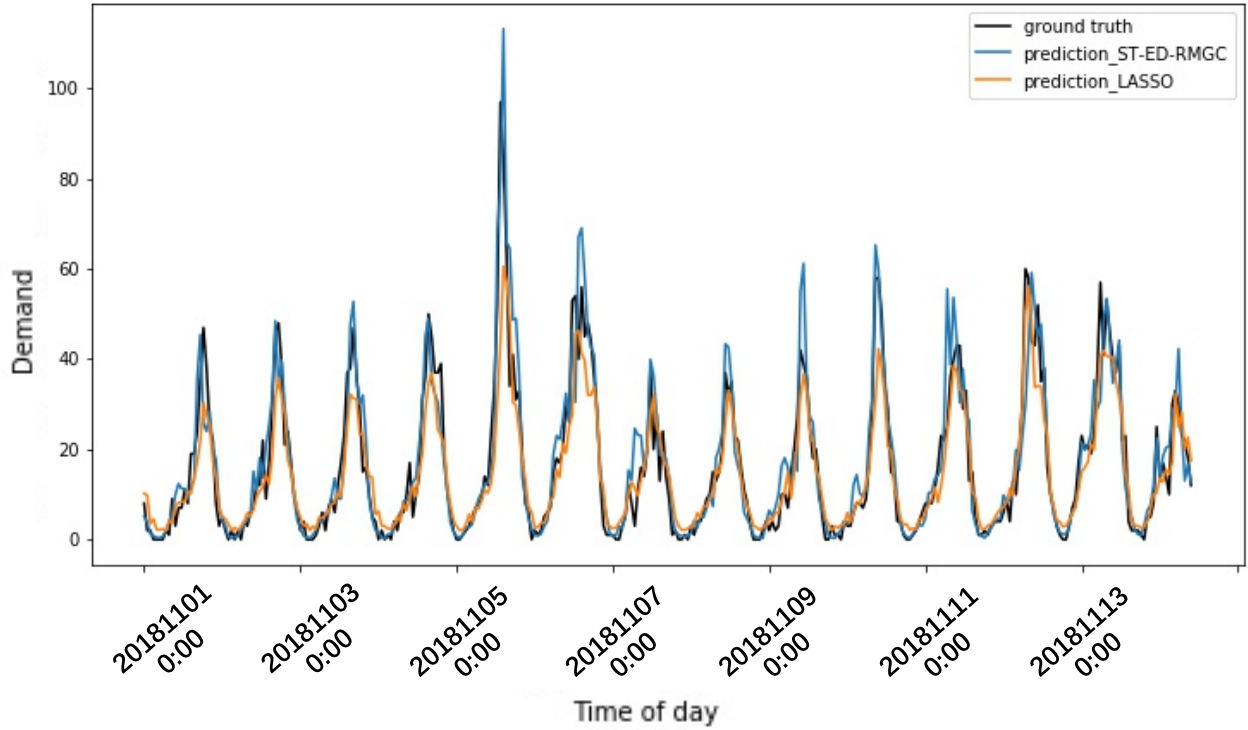

(a) OD pair A

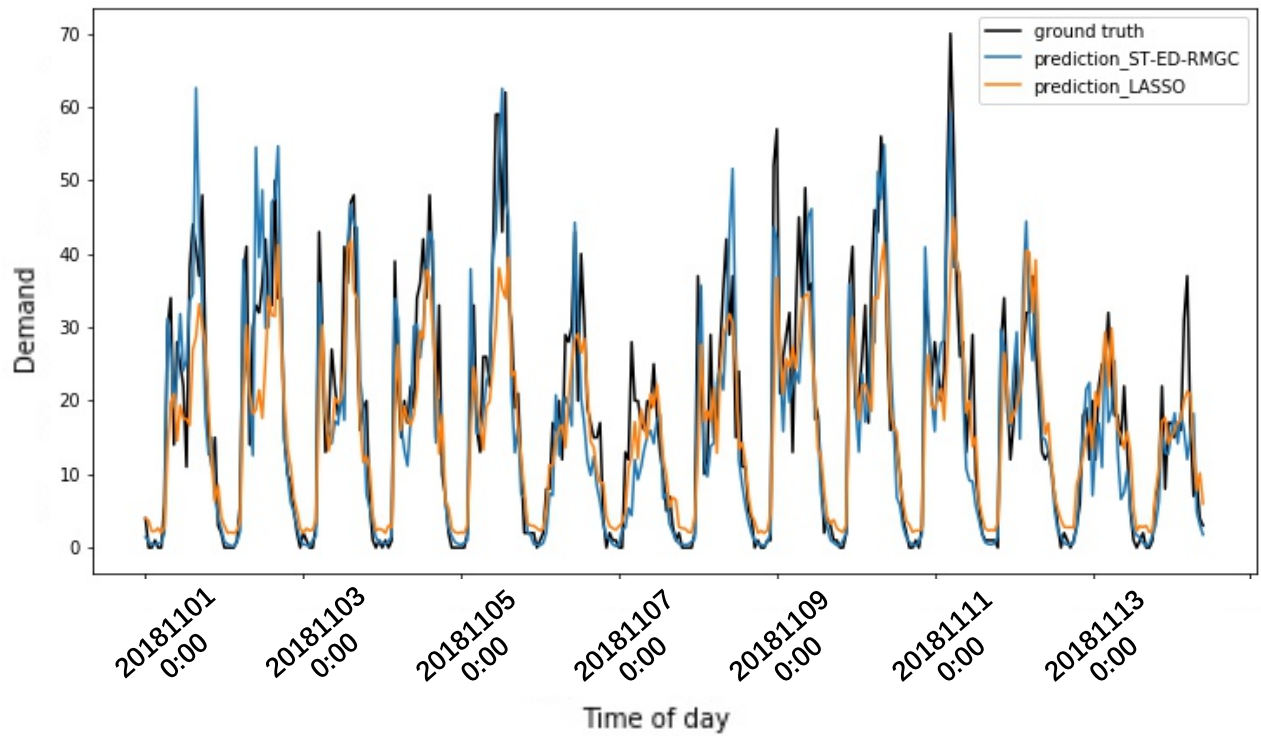

(b) OD pair B

**Figure 9.** The prediction result of different OD pairs over time

**6. Conclusions and Future Work**

In this paper, we study the OD-based ride-sourcing demand prediction problem. Compared with existing OD-based demand prediction approaches, we consider geographical and semantic correlations among OD pairs. That is, several OD graphs are constructed to measure the complex non-Euclidean spatio-temporal pair-wise relationships, from various aspects, including the geographical distances, neighboring relationships, mobility pattern correlations, and functional similarities. We then propose the ST-ED-RGN model with an encoder-decoder framework, which first encodes the spatial and temporal characteristics with RMGC networks and spatial LSTM networks into a latent vector space, and then utilizes RMGC networks again to decode the latent information for predicting the future OD demand. The proposed model is evaluated with real-world ride-sourcing mobility data in Manhattan, New York City, and is found to outperform the baselines by significant margins. For future work, we would like to involve more external features like weather, temperature, and emergencies to improve the predictive accuracy. Moreover, we propose to extend our model to predict the abnormal passenger demand due to accidents or public events.

**Acknowledgments**

The work described in this paper was supported by a grant from Hong Kong Research Grants Council under project HKUST16208619 and a NSFC/RGC Joint Research grant N_HKUST627/18. This work was also supported by the Hong Kong University of Science and Technology - DiDi Chuxing (HKUST-DiDi) Joint Laboratory.